\documentclass[manuscript]{aastex631}

\begin{document}

\title{Are Non-thermal Velocities in Active Region Coronal Loops Anisotropic?}

\correspondingauthor{Michael Hahn}
\email{mhahn@astro.columbia.edu}

\author{Michael Hahn}
\affiliation{Columbia Astrophysics Laboratory, Columbia University, 550 West 120th Street, New York, NY 10027}

\author{Mahboubeh Asgari-Targhi}
\affiliation{Harvard-smithsonian Center for Astrophysics, 60 Garden Street MS-15, Cambridge, MA 02138}

\author{Daniel Wolf Savin}
\affiliation{Columbia Astrophysics Laboratory, Columbia University, 550 West 120th Street, New York, NY 10027}

\begin{abstract}
	
We have measured line widths in active region coronal loops in order to determine whether the non-thermal broadening is anisotropic with respect to the magnetic field direction. These non-thermal velocities are caused by unresolved fluid motions. Our analysis method combines spectroscopic data and a magnetic field extrapolation. We analyzed spectra from the Extreme Ultraviolet Imaging Spectrometer on \textit{Hinode}. A differential emission measure analysis showed that many spectral lines that are commonly considered to be formed in the active region have a substantial contribution from the background quiet Sun. From these spectra we identified lines whose emission was dominated by the active region loops rather than background sources. Using these lines, we constructed maps of the non-thermal velocity. With data from the Helioseismic Magnetic Imager on the \textit{Solar Dynamics Observatory} and the Coronal Modeling System nonlinear force-free magnetic field reconstruction code, we traced several of the magnetic field lines through the active region. Comparing the spectroscopic and magnetic data, we looked for correlations of non-thermal velocity with the viewing angle between the line of sight and the magnetic field. We found that non-thermal velocities show a weak anti-correlation with the viewing angle. That is, the tendency is for the non-thermal velocity to be slightly larger in the parallel direction. This parallel broadening may be due to acoustic waves or unresolved parallel flows. 

\end{abstract}

\section{Introduction} \label{sec:intro}

Understanding coronal heating is a long-standing problem in solar physics, which is seen most dramatically in active regions, where the temperature and densities are signficantly greater than in the quiet Sun corona. Most theories agree that heating is ultimately driven by fluid motions in the photosphere. Theories differ, though, in how these motions transmit energy to the corona. Nanoflare theories consider the footpoint motions to drive a rearrangement of coronal field lines, which become stressed and then reconnect, releasing energy and heating the corona. Wave-turbulence theories consider the fluid motions to drive waves that travel into the corona, where they drive turbulence and heating. Unfortunately, many commonly used diagnostics for coronal heating, such as temperature, density, differential emission measure (DEM) distribution, intensity time series, etc. are not able to clearly distinguish between these mechanisms \citep[][]{Klimchuk:PTA:2015}. 

Measurements of the non-thermal velocity $v_{\mathrm{nt}}$ within the active region coronal loops might provide a new diagnostic for the heating process. These velocities quantify the Doppler broadening of spectral lines, which is caused by unresolved fluid motions. Of particular value, is the anisotropy of the non-thermal velocity, that is, the components parallel $v_{\mathrm{nt}, \parallel}$ and perpendicular $v_{\mathrm{nt}, \perp}$ to the mean magnetic field. The utility is that $v_{\mathrm{nt}, \perp}$ is directly related to wave turbulence, as Alfv\'en waves cause transverse fluid motions with $v_{\mathrm{nt},\perp}$ proportional to the wave amplitude. The parallel component may be sensitive to any heating, whether by waves or nanoflares, as localized heating will drive flows parallel to the loop increasing $v_{\mathrm{nt},\parallel}$. 

Only a few existing measurements have attempted to measure the anisotropy of $v_{\mathrm{nt}}$, but the results are ambiguous. \citet{Hara:ApJ:1999} observed loops at the solar limb and characterized them as being either face-on or edge-on. For face-on loops the line of sight is perpendicular to the magnetic field everywhere, while for edge-on loops the line of sight is perpendicular to the loop near the base and parallel near the loop top. Based on their measurements, \citet{Hara:ApJ:1999} concluded that line widths were $3$--$5$~$\mathrm{km\,s^{-1}}$ broader in the perpendicular direction than in the parallel direction, so $v_{\mathrm{nt},\perp} > v_{\mathrm{nt}, \parallel}$. 

A different study appears to have found the opposite result. \citet{Hara:ApJ:2008} measured the center-to-limb variation of $v_{\mathrm{nt}}$ for several lines. The relevance to anisotropy is that near the limb the line of sight tends to be perpendicular to the magnetic field, whereas at disk center the view tends to be along the field. \citet{Hara:ApJ:2008} found broadly overlapping distributions of $v_{\mathrm{nt}}$ at both locations, with a slight tendency for $v_{\mathrm{nt}}$ to be larger at disk center. This could be interpreted as weak evidence for $v_{\mathrm{nt},\parallel} > v_{\mathrm{nt}, \perp}$. Other studies of center-to-limb variation have not found indications of anisotropy \citep[][]{DelZanna:LRSP:2018}. 

A limitation of these measurements has been that the angle between the magnetic field and the line of sight is very uncertain. To overcome this limitation, we have used a nonlinear force free magnetic field model to extrapolate the magnetic field of the loops and specify their orientation. With this information, we can resolve the uncertainty in the angle between the line of sight and the magnetic field. 

Our objective is to determine whether there is a relationship between the inclination angle of the magnetic field to the line of sight and $v_{\mathrm{nt}}$. For our analysis we use observations from \textit{Hinode} and the \textit{Solar Dynamics Observatory (SDO)}, which are described in Section~\ref{sec:obs}. The analysis of these data to determine the line widths and extrapolate the magnetic field is given in Section~\ref{sec:anal}. Our results are presented in Section~\ref{sec:res}, along with a discussion of various systematic uncertainties and the interpretation of the results. We present our conclusions in Section~\ref{sec:conclusion}. 

\section{Instrument and Observations}\label{sec:obs}

We analyzed an active region observed by the \textit{Hinode} Extreme Ultraviolet Imaging Spectrometer \citep[EIS;][]{Culhane:SolPhys:2007} on 2011-04-19 at 12:03:27 UT. For this observation the 1$^{\prime\prime}$ slit of EIS was rastered across the active region spanning from about $x=-101^{\prime\prime}$ to $118^{\prime\prime}$ in the horizontal direction and from $y=107^{\prime\prime}$ to $618^{\prime\prime}$ in the vertical direction. The exposure time for each raster position was about $60$~s, so that the data were collected over a total of about two hours. The raw data were prepared for further analysis using standard EIS calibration routines to correct for dark current, cosmic rays, and offsets in the wavelength and spatial scales. 

The radiometric calibration of EIS has been identified as a potential source of systematic error for line width studies \citep{Brooks:ApJ:2016, Testa:ApJ:2016}. This is because over time the EIS sensitivity, parameterized as an effective area, has changed non-uniformly versus wavelength. In order to account for these issues, the line shape analysis described here has been performed without applying the absolute calibration. However, for the DEM analysis, the absolute calibration is needed and for those results we do apply the effective area calibration as updated by \citet{Warren:ApJS:2014} to account for the wavelength-dependent degradation over time. 

As a test, for a few isolated lines we compared the total line widths from the data with and without the absolute calibration. The line widths using the uncalibrated data are smaller by $< 1\%$ than those based on the calibrated data. For example, the median line width for Fe~\textsc{xv} 284.16~\AA\ was about 0.0002~\AA\ smaller in the uncalibrated data, which had a median line width was $0.0347$~\AA. This corresponds to a difference in the median line width of less than 0.2~$\mathrm{km\,s^{-1}}$. This may also be compared to the typical uncertainty in an individual fit for the Fe~\textsc{xv} line width, which was about 0.002~\AA. So, the systematic difference due to the calibration is only about 10\% of the typical fitting uncertainty. Based on these tests, any systematic uncertainties in the line widths due to the absolute calibration are likely negligible. For our analysis we follow \citet{Brooks:ApJ:2016} and derive line widths from the uncalibrated data. 

The magnetic field reconstruction is based on magnetogram data from the Helioseismic Magnetic Imager \citep[HMI;][]{Scherrer:SolPhys:2012} on \textit{SDO}. 
We also compare the reconstruction to images from the Atmospheric Imaging Assembly \citep[AIA;][]{Lemen:SolPhys:2012} on \textit{SDO} in order to verify that the extrapolated magnetic fields resemble the observed magnetic loops. Additionally, we have compared the AIA data to the EIS data in order to co-align the magnetic field model with the EIS spectra. 

\section{Analysis}\label{sec:anal}

Before describing the process in detail, we give a brief outline of the steps in the analysis: We first identified lines dominated by emission from the active region loops, i.e., with minimal contamination from quiet Sun in the foreground or background. For those lines, we subtracted the instrumental broadening and estimated the thermal velocity to extract the non-thermal velocity, $v_{\mathrm{nt}}$, and constructed a map of $v_{\mathrm{nt}}$ in the observed region. Next, we used a nonlinear force-free field model to extrapolate the photospheric magnetic field into the corona and trace out the active region loops. We traced $v_{\mathrm{nt}}$ along these field lines and determined the correlation between $v_{\mathrm{nt}}$ and the angle between the magnetic field and the line of sight. 

\subsection{Spectroscopic Data}

All of the EIS data, both calibrated and uncalibrated, were fit with a Gaussian function in order to measure their centroid, line width, and intensity. The main quantity we are interested in is $v_{\mathrm{nt}}$, which is derived from the line width. We are interested only in $v_{\mathrm{nt}}$ from the active region coronal loops. As the corona is optically thin, the emission for all of the lines is distributed along the line of sight and contains emission from the quiet Sun. In order to avoid lines with a significant quiet Sun contribution, we identified lines whose emission is mainly coming from the active region. 

A common method to restrict the analysis to emission from the active region is to select lines based on their formation temperature. The justification for this is that such hot lines must be emitted from the active region. However even relatively hot lines can have significant quiet Sun contributions. For example, Fe~\textsc{xiv} is formed at $\log T_{\mathrm{f}} = 6.3$ (here and throughout, temperatures are in K) and is often considered an active region line; but we found that only $10\%$ of the intensity of this line comes from active region temperatures and the rest is emitted in the quiet Sun. 

A more systematic way to constrain the emitting structure is based on a DEM analysis. The DEM, $\phi(T_{\mathrm{e}})$, describes the distribution of material along the line of sight as a function of electron temperature, $T_{\mathrm{e}}$. The DEM is related to the line intensity of the transition from level $j$ to level $i$ by 
\begin{equation}
I_{ji} = \frac{1}{4\pi}\int{ G_{ji}(T_{\mathrm{e}})\phi(T_{\mathrm{e}}) dT_{\mathrm{e}}}. 
\label{eq:demdef}
\end{equation}
Here, $G_{ji}(T_{\mathrm{e}})$ is the contribution function and describes the level populations, ionization balance, elemental abundance, and radiative decay rates. The integral is over all temperatures. The needed atomic data are tabulated by the CHIANTI atomic database \citep{Dere:AA:1997, DelZanna:ApJ:2021}. The DEM can be derived from a set of measured line intensities by inverting Equation~(\ref{eq:demdef}). Once the DEM is found, the intensity due to emission from material within a given temperature range can be found by integrating Equation~(\ref{eq:demdef}) over the range of interest. For example, one could define a temperature-based criterion for emission to be from the active region as having the intensity predominantly from material above $\log T_{\mathrm{e}} = 6.3$. 

To simplify the DEM analysis and the identification of emission sources, we have assumed a relatively simple functional form for the DEM. Our model function assumes that there are several discrete structures along the line of sight corresponding roughly to the transition region, coronal holes, quiet Sun, and active regions. Each of these structures is represented in our parameterization by a delta function with an emission measure amplitude $\mathrm{EM}_{k}$ and a temperature $T_{k}$, where the temperatures for the $T_{k}$ are assumed to fall within a characteristic range. The total DEM is the sum of the contribution from these structures: 
\begin{equation}
\phi(T_{\mathrm{e}}) = \sum_k \mathrm{EM}_{k}\delta(T_{\mathrm{e}} - T_{k}). 
\label{eq:demodel}
\end{equation}

Although it appears very simple, this function is detailed enough to capture the thermal structure of the emission to the extent permitted by the data. Using a Bayesian analysis,  \citet{Dere:ApJ:2022} has shown that observational and atomic physics uncertainties limit the extent to which the DEM can be constrained. Specifically, \citeauthor{Dere:ApJ:2022} concluded that 4 temperature and emission measure pairs were the maximum number of parameters that could be fit without introducing additional assumptions. For example, many emission measure analysis methods make assumptions that the temperature structure should be smooth. 

DEM analyses of the solar atmosphere typically show peaks at certain discrete temperatures. \citet{Feldman:PoP:2008} argued that DEMs can be considered as having several isothermal components at temperatures of about $\log T_{\mathrm{e}} \approx 5.75$, $5.95$, $6.15$, and $6.50$, which they identified with emission from the transition region, coronal hole, quiet Sun, and active region, respectively. \citet{Dere:ApJ:2022} found very similar peaks using a different method and with twenty years of improved atomic data. Our simple model is, thus, well supported by the observed thermal structure. In any case, for our purposes we need not be concerned whether our discontinous DEM or a very smooth one is more representative of the real Sun. It suffices that it is representative enough to serve as a criterion for separating the active region emission from background sources. 

A practical advantage of this DEM model is that it simplifies much of the analysis. Using the functional form of Equation~(\ref{eq:demodel}) in Equation~(\ref{eq:demdef}), we can find the DEM by inverting 
\begin{equation}
I_{ji} = \sum_{k} \mathrm{EM}_{k}G_{ji}(T_{k}). 
\label{eq:deminv}
\end{equation}
The inversion was performed using a least squares fitting algorithm. The free parameters in our model are four $\mathrm{EM}_k$ and $T_{k}$ pairs. In practice, we have performed the fit using $\log(\mathrm{EM}_k)$ and $\log T_{k}$, since the parameters can vary over orders of magnitude. We consider the active region to be the highest temperature component with $\log T_{\mathrm{e}} > 6.3$. Denoting this component as the one with $k=\mathrm{AR}$, we can easily find the fraction of emission for a particular line that comes from the active region using 
\begin{equation}
f_{\mathrm{AR},ji} = \frac{\mathrm{EM_{AR}} G_{ji}(T_{\mathrm{AR}})}{\sum_k \mathrm{EM}_{k} G_{ji}(T_{k})}. 
\label{eq:farji}
\end{equation}
It is also useful to define a quantity to denote the fraction of the total emission measure coming from the active region, which is given by 
\begin{equation}
f_{\mathrm{AR},ji} = \frac{\mathrm{EM_{AR}}}{\sum_k \mathrm{EM}_{k}}. 
\label{eq:far}
\end{equation}
A similar definition could be made for the quiet Sun contribution or the other components. 

This analysis shows that many emission lines that are commonly considered to be formed in the active region are really mainly formed in the quiet Sun. For such lines, the large amount of quiet Sun material compared to the thin active region loop compensates for a reduced value of $G(T_{\mathrm{e}})$ in the quiet Sun. For iron ions, we found that the active region contributed about $10\%$ of the emission for Fe~\textsc{xiv} 270.52~\AA\ ($\log T_{f} = 6.3$), 50\% for Fe~\textsc{xv} 284.16~\AA\ ($\log T_{f} = 6.35$), and 90\% for Fe~\textsc{xvi} 262.98~\AA\ ($\log T_{f} = 6.45$). Of these, only the Fe~\textsc{xvi} line emission comes predominantly from the active region. Other suitable lines we found were Ca~\textsc{xv} 201.00~\AA\ ($\log T_{f} = 6.65$) with $f_{\mathrm{AR},ji} \approx 100\%$ and Ar~\textsc{xiv} 187.96~\AA\ ($\log T_{f} = 6.55$) with $f_{\mathrm{AR},ji} = 99\%$. 

Gaussian profiles were fit to each spectral line in order to determine the line width $\Delta \lambda$. Then the instrumental broadening, $\Delta \lambda_{\mathrm{Inst}}$, was subtracted from the total measured line width. Calibrations of  $\Delta \lambda_{\mathrm{Inst}}$ have been given by \citet{Young:EIS:2011} and by \citet{Hara:ApJ:2011}. We have used the calibrations from \citet{Hara:ApJ:2011}, which were obtained by cross calibrating EIS line widths with measurements from ground-based instruments. 

After subtracting the instrumental width, the remaining line width comes from physical effects. It is convenient to describe this line width as an effective velocity, defined by 
\begin{equation}
v_{\mathrm{eff}} \equiv \sqrt{\left(\frac{2 k_{\mathrm{B}}T_{\mathrm{i}}}{M} + v_{\mathrm{nt}}^{2}\right)} = \sqrt{v_{\mathrm{th}}^2 + v_{\mathrm{nt}}^2}.
\label{eq:defVeff}
\end{equation}
In our analysis, $v_{\mathrm{eff}}$ comes from the Gaussian fits to the spectral lines corrected for the instrumental broadening. In order to find $v_{\mathrm{nt}}$ we must subtract the thermal velocity, $v_{\mathrm{th}}$, which depends on the ion temperature, $T_{\mathrm{i}}$, and the ion mass $M$. 

For the active region, we have assumed that $T_{\mathrm{i}} = T_{\mathrm{e}}$, because the active region loops have a relatively high density $\approx 10^9$~$\mathrm{cm^{-3}}$, so that collisions between electrons and ions can maintain thermal equilibrium. As we focus on the active region lines, we take this temperature to be the $T_{\mathrm{AR}}$ parameter from our DEM analysis using Equation~(\ref{eq:demodel}). 

The assumption that $T_{\mathrm{i}} = T_{\mathrm{e}}$ is also plausible based on the measured $v_{\mathrm{eff}}$. In the pixels where the active region part of the DEM accounts for more than half of the total emission, that is $f_{\mathrm{AR}} > 0.5$, we have found that the average values for $v_{\mathrm{eff}}$ are $33.9 \pm 9.2$~$\mathrm{km\,s^{-1}}$ for Fe~\textsc{xvi}, $42 \pm 29$~$\mathrm{km\,s^{-1}}$ for Ar~\textsc{xiv}, and $51 \pm 23$~$\mathrm{km\,s^{-1}}$ for Ca~\textsc{xv}. Here, the uncertainty represents the standard deviation among the values from each pixel and not the propagated uncertainty from the Gaussian fits. To test whether the data are consistent with $T_{\mathrm{i}} = T_{\mathrm{e}}$ we considered a less restrictive assumption: that $T_{\mathrm{i}}$ is the same for all the ions, but may differ from the $T_{\mathrm{e}}$. This is less restrictive, because of the large mass difference between ions and electrons, which causes collisions to couple the ion temperatures more strongly to one another and to the protons than to the electrons. Under this assumption, we can perform a fit to Equation~(\ref{eq:defVeff}) with the ion mass $M$ as the dependent variable and $T_{\mathrm{i}}$ and $v_{\mathrm{nt}}$ as free parameters. A least squares fit finds that $\log T_{\mathrm{i}} = 6.6 \pm 4.0$. This average value matches the $T_{\mathrm{e}}$ expected for an active region. Clearly there is a huge uncertainty, which is partly due to the spread in $v_{\mathrm{eff}}$ throughout the active region, but more importantly to the extremely limited range of masses for these three ions. In particular, Ca and Ar have nearly the same ion mass. 

In order to account for the uncertainties in $v_{\mathrm{eff}}$ and $T_{\mathrm{AR}}$ when obtaining $v_{\mathrm{nt}}$, we have used a Monte-Carlo type approach. We drew normally distributed random numbers with means equal to the values inferred from the measurements of $v_{\mathrm{eff}}$, and $\log T_{\mathrm{AR}}$, and standard deviations equal to the uncertanities on those values. These distributions sample from probability distributions for each of these quantities. We computed $v_{\mathrm{nt}}$ for each sampled pair using Equation~(\ref{eq:defVeff}). About 5000 samples were used to derive the probability distribution for $v_{\mathrm{nt}}$. In some cases a sampled pair of line width and temperature produces an unphysical value with $v_{\mathrm{nt}} < 0$. Such values were ignored based on the prior assumption that the probability of a negative non-thermal velocity is zero. In order to represent $v_{\mathrm{nt}}$ at each location with a single value and an uncertainty, we have used the median of the positive $v_{\mathrm{nt}}$ distribution. For the uncertainty, we computed the range about the median that  would contain 68.2\% of the probability distribution, analogous to the $1\sigma$ range. The range was roughly symmetric above and below the median, so we take half that range as an estimate of the $1\sigma$ uncertainty. 

One issue with this method is that it can appear to inflate the inferred value of $v_{\mathrm{nt}}$ at locations where there are large uncertainties. This is because $v_{\mathrm{nt}}$ is bounded below by zero, but there is no upper bound. When the uncertainties in $v_{\mathrm{eff}}$ or $T_{\mathrm{AR}}$ are large, the probability distribution for $v_{\mathrm{nt}}$ is relatively flat and extends over a broad range. As a result, the median value for $v_{\mathrm{nt}}$ can become very large and proportional to the uncertainties. In order to mitigate this problem, we have limited our analysis to locations deep in the active region where the high temperature emission is strongest, as discussed in Section~\ref{sec:res}. 

\subsection{Magnetic Field Reconstruction}

The angle $\theta$ between the line of sight and the magnetic field of the loop is found from magnetic field extrapolation, which is performed using the Coronal Modeling System \citep[CMS;][]{vanB2004, Asgari:ApJ:2012, Asgari:ApJ:2021}. We used line-of-sight magnetogram data from HMI, which provides full-disk magnetograms with 0.5$^{\prime\prime}$ resolution and a 45~s cadence \citep[][]{Scherrer:SolPhys:2012}. Our CMS-modeled magnetic field is also constrained by extreme ultraviolet (EUV) images. These data are taken from AIA \citep[][]{Lemen:SolPhys:2012}. 

To extrapolate the coronal magnetic field, CMS inserts a magnetic flux rope into a potential-field model of the target region and then applies magneto-frictional relaxation \citep{Yang1986, vanB2000}. The calculation is performed on a high-resolution domain surrounding the target region and a global potential field model is used to set the boundary conditions. The method is flexible and accounts for the spherical geometry of the Sun. The results also provide information about the stability of the magnetic fields. The axial and poloidal magnetic flux of the inserted flux rope are free parameters with respect to the magnetogram. These parameters are constrained by matching the extrapolated magnetic fields to the AIA EUV images. As a result, CMS provides accurate orientations of the coronal magnetic fields as assessed by inspection of the field line fits to the observed coronal loops. CMS has been successfully applied to studies of filaments \citep{vanB2004}, active regions \citep{Bobra2008, Su2009a, Su2009b, Su2011}, X-ray sigmoids \citep[][]{Sav2009}, and other loops \citep{Asgari:ApJ:2012, Asgari:ApJ:2013, Asgari:ApJ:2014}.

\begin{figure}
	\centering \includegraphics[width=0.9\textwidth]{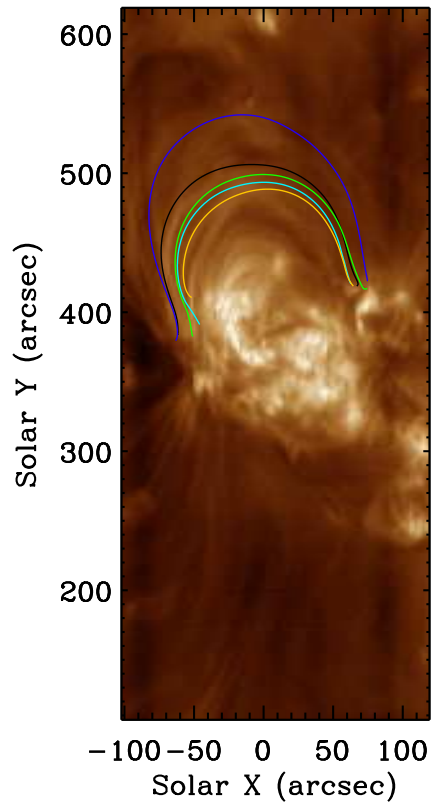}
	\caption{\label{fig:fe12image} Image of the intensity of the Fe~\textsc{xii} 195.12~\AA\ line measured by EIS for this observation. The colored curves show the selected magnetic field lines traced by the CMS magnetic field extrapolation superimposed on the EIS data. 
	}
\end{figure}

The magnetic field reconstructions were co-aligned to the spectra by comparing the EIS the Fe~\textsc{xii} 195.12~\AA\ line emission with the 193~\AA\ channel of AIA, for which the emission is primarily from the same ion. Once co-aligned, we could project the magnetic field lines onto the EIS data. Figure~\ref{fig:fe12image} shows selected magnetic field lines computed by CMS superimposed on an image in the Fe~\textsc{xii} 195.12~\AA\ line measured by EIS. Fe~\textsc{xii} formation peaks at relatively cool temperatures ($\log T_{\mathrm{e}} = 6.2$), but its contribution function extends into the active region temperature range and the emission from the denser active region loops enhances the intensity above the background level, rendering the loops visible. 

The CMS reconstruction provides the three-dimensional vector magnetic field throughout the corona. Magnetic field lines, such as those in Figure~\ref{fig:fe12image} are traced out using this model. Thus, CMS provides the coordinates of the magnetic field lines and the vector magnetic field at each point along the traced field lines. For each spatial pixel in the EIS spectra along the loop, we found the angle between the magnetic field and the line of sight. In the CMS coordinate system the $\hat{z}$ direction is the line of sight, so the angle between the magnetic field and the line of sight is $\theta = \cos^{-1}(B_{z}/B)$, where $B$ is the magnitude of the magnetic field strength.

\section{Results and Discussion}\label{sec:res}

\begin{figure}
	\centering \includegraphics[width=0.9\textwidth]{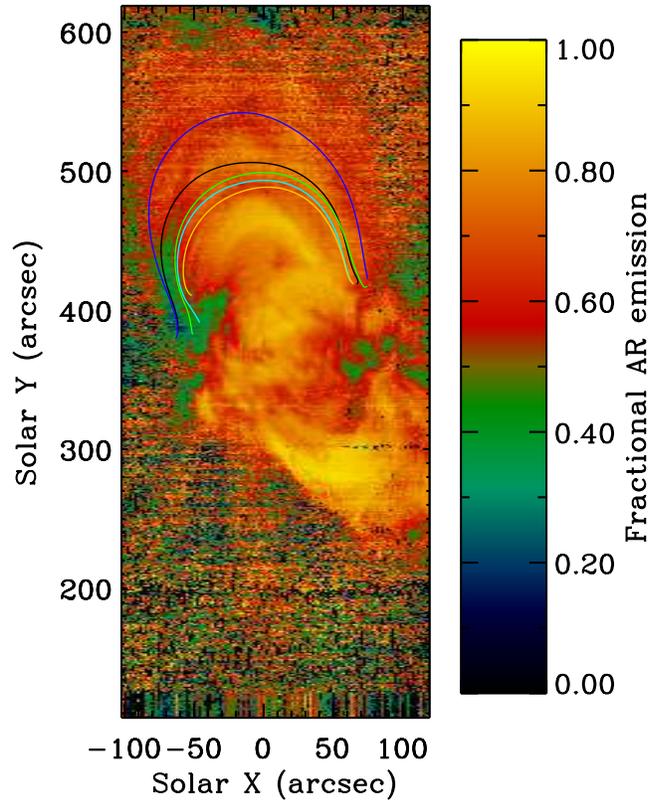}
	\caption{\label{fig:arfrac} Map of $f_{\mathrm{AR}}$ throughout the EIS observation, showing fraction of the emission measure in each pixel that comes from the active region component of the DEM. The colored curves show the selected magnetic field lines from Figure~\ref{fig:fe12image} superimposed on the EIS data. 
}
\end{figure}

We looked for correlations between $v_{\mathrm{nt}}$, $v_{\mathrm{eff}}$, and $\theta$ for all the points along the five traced coronal loops. One systematic correlation that we need to control for is the influence of noise in the measurements of $v_{\mathrm{eff}}$. As discussed above, we were focusing on lines from Fe~\textsc{xvi}, Ca~\textsc{xv}, and Ar~\textsc{xiv}, because their intensity is dominated by active region plasma. Outside of the active region, these lines are very weak and the line widths have a large uncertainty. As discussed above, since $v_{\mathrm{nt}}$ cannot be less than zero, the uncertainty propagation will tend to increase the statistical median value of the derived $v_{\mathrm{nt}}$ in regions where the lines are weak. In order to mitigate these correlations, we limit the analysis to regions where the active region contributes at least half of the total emission measure $f_{\mathrm{AR}} > 0.5$. Figure~\ref{fig:arfrac} maps $f_{\mathrm{AR}}$ in this region. 

\begin{figure}
	\centering \includegraphics[width=0.9\textwidth]{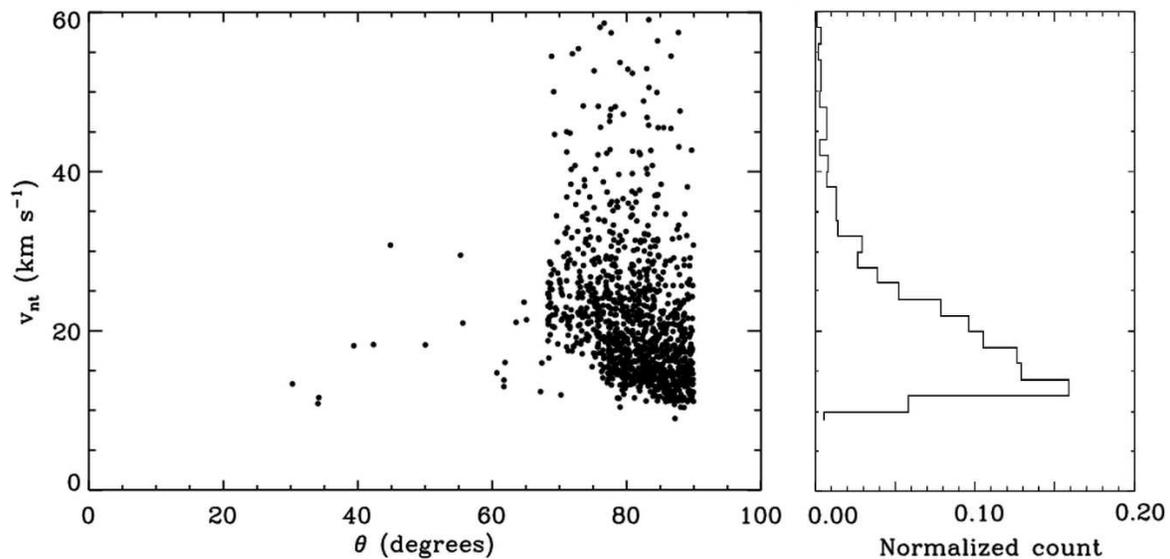}
	\caption{\label{fig:Fe16ani} Composite of $v_{\mathrm{nt}}$ versus $\theta$ for Fe~\textsc{xvi} points drawn from all of the traced loops that satisfy the criteria that $f_{\mathrm{AR}} \geq 0.5$. Because of the zero lower bound, large $v_{\mathrm{nt}}$ values can occur at locations where there is more uncertainty in $v_{\mathrm{eff}}$ or $T_{\mathrm{AR}}$. The histogram on the right helps to illustrate the number of such outliers compared to the bulk of the $v_{\mathrm{nt}}$ distribution. 
	}
\end{figure}

Figure~\ref{fig:Fe16ani} illustrates the relationship between $v_{\mathrm{nt}}$ and $\theta$ for the Fe~\textsc{xvi} 262.98~\AA\ line. For this line, the Spearman correlation coefficient, $\rho_c$, was statistically significant with $\rho_c = -0.35$. The analysis based on Ca~\textsc{xv} 201.00~\AA\ agreed with $\rho_c = -0.29$. We also performed the analysis for the Ar~\textsc{xiv} 187.96~\AA\ line, which showed a weaker correlation of $\rho_c = -0.14$. %
The correlation is likely weaker because the Ar~\textsc{xiv} line has a lower intensity than either of the other two lines and the resulting uncertainty in the measured line widths obscures the correlation. For all three emission lines, the correlations and significance levels remained about the same if we raised the threshold criterion to $f_{\mathrm{AR}} = 0.6$. Overall, there is a weak tendency for $v_{\mathrm{nt}}$ to be greater in the direction parallel to the loop magnetic field. 

We have also looked for correlations between $v_{\mathrm{eff}}$ and $\theta$. Doing so may obviate any systematic uncertainties introduced by the subtraction of the thermal width, but at the expense of making correlations by not correcting for the temperature dependence. We found that the correlations with $v_{\mathrm{eff}}$ tend to be weaker and less significant than with $v_{\mathrm{nt}}$ itself. For Fe~\textsc{xvi}, the Spearman correlation coefficient $v_{\mathrm{eff}}$ and $\theta$ is not statistically significant when the threshold is $f_{\mathrm{AR}} > 0.5$. With a more restrictive criterion of $f_{\mathrm{AR}} > 0.6$, we find a correlation coefficient of -0.36, which is comparable to what we find for $v_{\mathrm{nt}}$. For Ca~\textsc{xv}, $v_{\mathrm{eff}}$ shows a weak inverse correlation with $\theta$ having $\rho_c=-0.133$. The Ar~\textsc{xiv} data showed no significant correlations. So, we find that for $v_{\mathrm{eff}}$ the inverse correlation between $v_{\mathrm{eff}}$ and $\theta$ probably exists, but is weaker than for $v_{\mathrm{nt}}$.

We can readily estimate the magnitude of the perpendicular $v_{\mathrm{nt},\perp}$ by taking the average $v_{\mathrm{nt}}$ value for those pixels with large angles. Here we focus on the Fe~\textsc{xvi} line, since it is the brightest line with the smallest uncertainties. For the range $\theta = 80$--$90^{\circ}$, we find that the average $v_{\mathrm{nt},\perp} = 22 \pm 11$~$\mathrm{km\,s^{-1}}$. It is more difficult to estimate the parallel $v_{\mathrm{nt},\parallel}$ since there are very few lines of sight in this observation that view the loop at a small nearly parallel angle. One possibility is to extrapolate to small angles by performing a fit to the data using \citep{Hahn:ApJ:2013}: 
\begin{equation}
v_{\mathrm{nt}} = \sqrt{ v_{\mathrm{nt, \parallel}}^2 \cos^2{\theta} + v_{\mathrm{nt}, \perp}^2 \sin^2{\theta}}. 
\label{eq:fitcomps}
\end{equation}
Such a fit yields $v_{\mathrm{nt},\parallel} = 28 \pm 3$~$\mathrm{km\,s^{-1}}$ and $v_{\mathrm{nt},\perp} = 19.0 \pm 0.3$~$\mathrm{km\,s^{-1}}$. These uncertainties are probably underestimates as they are weighted by the uncertainties of each pixel, but ignore possibly real pixel-to-pixel variations. Moreover, due to the zero lower bound on $v_{\mathrm{nt}}$ discussed above, these magnitudes are likely systematically overestimated. 

We can estimate the energy flux in the loop due to the parallel and perpendicular non-thermal velocities if we interpret them as the amplitudes of sound waves and Alfv\'en waves, respectively. In that case, the energy flux of the waves is approximately $F \approx \rho v_{\mathrm{nt}}^2 V_{g}$, where $\rho$ is the mass density and $V_{g}$ is the appropriate group velocity of either the sound speed or the Alfv\'en speed, respectively. Given the temperatures, densities, and magnetic fields of these loops the sound speed is roughly $c_{\mathrm{s}}\approx 300$~$\mathrm{km\,s^{-1}}$ and the Alfv\'en speed is $V_{\mathrm{A}}\approx 1000$~$\mathrm{km\,s^{-1}}$. Using the above non-thermal velocities and an estimated density of $10^{9}$~$\mathrm{cm^{-3}}$, the energy fluxes for sound waves would be about $F_{\mathrm{s}} \approx 400$~$\mathrm{W\,m^{-2}}$ and for Alfv\'en waves $F_{\mathrm{s}} \approx 600$~$\mathrm{W\,m^{-2}}$. This suggests there is a similar amount of energy in the parallel and perpendicular fluctuations. 

It is implicit in the above analysis that $v_{\mathrm{nt}}$ and its components do not vary along the loop. This assumption is unavoidable as there are not enough measurables to extract possible actual variations in $v_{\mathrm{nt, \perp}}$, $v_{\mathrm{nt},\parallel}$, and $T_{\mathrm{i}}$ simultaneously. This assumption is also made in other studies, such as \citet{Hara:ApJ:1999} and \citet{Hara:ApJ:2008} mentioned in the introduction. Here, we have investigated some factors that might limit the validity of this assumption by examining how the parameters of interest vary with distance along the loop. 
	
\begin{figure}
	\centering \includegraphics[width=0.9\textwidth]{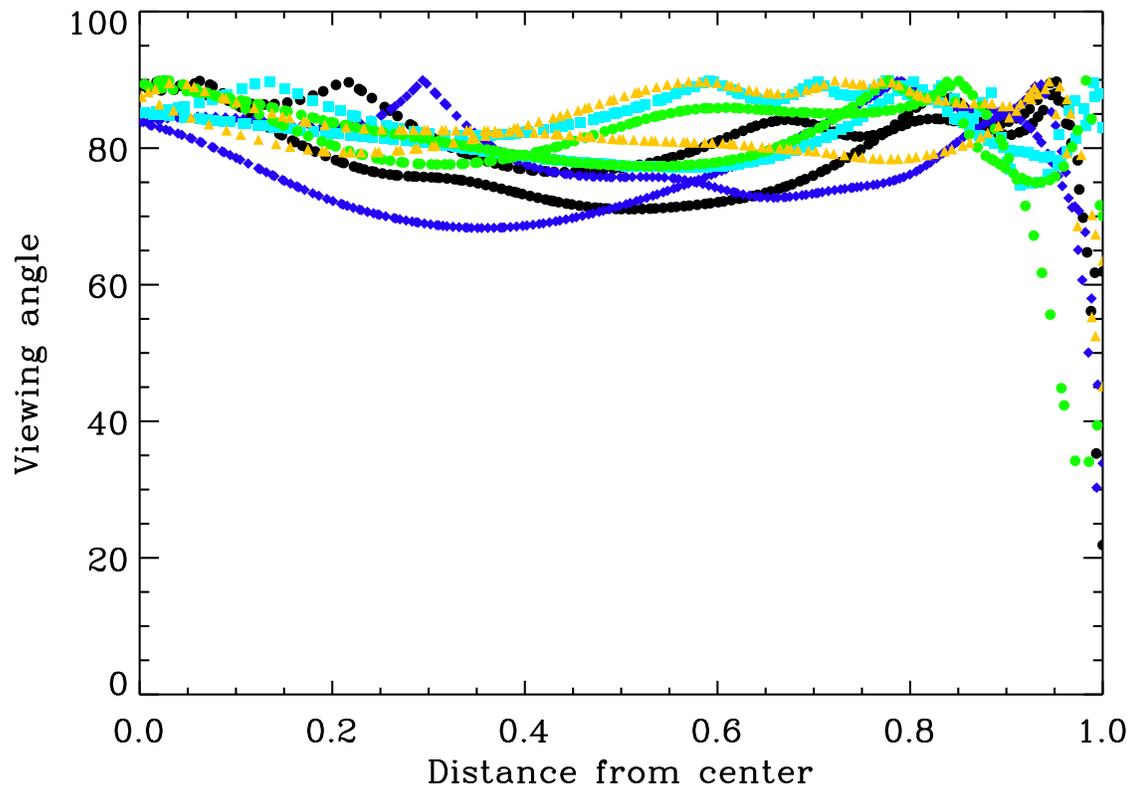}
	\caption{\label{fig:distangle} Angle $\theta$ between the line of sight and the loop magnetic field as a function of the normalized distance from the loop center for the five traced loops. The different symbols and colors indicate different loops. The colors match those in Figure~\ref{fig:fe12image}. 
	}
\end{figure}

For this observation, we are looking down onto the loop so that the viewing angle is most parallel near the footpoints. Because of this perspective, the legs of the loop are greatly foreshortened and cover only a few pixels. This is illustrated in Figure~\ref{fig:distangle}, where we have plotted the viewing angle $\theta$ as a function distance from the center of the loop normalized by the loop half length, $s$. That is, the center of the loop is at $s=0$ and the footpoints are at $s=1$. 

The temperature, $T_{\mathrm{AR}}$ has an inverse correlation with $s$, though it is quite weak with a correlation coefficient of only $-0.15$. A decreasing temperature toward the footpoints is theoretically expected for active region loops, \citep[e.g.,][]{Rosner:ApJ:1978}. One possible reason for the weak correlation here is our assumption that lower temperatures are due to quiet Sun contamination. It is possible that we are misinterpreting some positions near the loop footpoints as reflecting quiet Sun background, when those positions are actually viewing emission from the cooler part of the loop. 

One might be concerned whether the subtraction of the thermal width from $v_{\mathrm{eff}}$ could introduce a systematic effect that would bias the $v_{\mathrm{nt}}$ data due to the above correlation between temperature and $s$. Since a weak anisotropy is observed also in $v_{\mathrm{eff}}$, the stronger correlation with $v_{\mathrm{nt}}$ suggests that $v_{\mathrm{nt}}$ really is varying and we have subtracted the thermal width that was obscuring that correlation when looking at $v_{\mathrm{eff}}$. 
	
It is possible that the measured correlation with $v_{\mathrm{nt}}$ is not caused by anisotropy directly, but rather that the apparent correlation between $v_{\mathrm{nt}}$ and $\theta$ could be an artifact of an underlying real relationship with distance along the loop from a footpoint. However, the relationship between $s$ and $\theta$ is not monotonic (see Figure~\ref{fig:distangle}). Rather, $\theta$ varies significantly within the range of about $60$--$90^{\circ}$ over a large fraction of the loop length, and small angles are only seen very close to the footpoints. Consequently, the observed correlations of $v_{\mathrm{nt}}$ with angle are mostly decoupled from any variations with $s$; though there was a weak, but significant, correlation between $v_{\mathrm{nt}}$ and $s$ with correlation coefficient $-0.11$ based on the Fe~\textsc{xvi} data. This correlation is weaker than that between $v_{\mathrm{nt}}$ and $\theta$, so it is likely that the observed correlations are mainly due to anisotropy. 

\section{Conclusion}\label{sec:conclusion}

We have measured the variation of $v_{\mathrm{nt}}$ along active region coronal loops and compared those measurements to the viewing angle of the line of sight with respect to the magnetic field direction, $\theta$. This analysis is made possible by a nonlinear force-free field extrapolation to infer the magnetic field direction. Additionally, we found that background emission from the quiet Sun is strong, so that only a few lines with very high formation temperature are actually dominated by the active region. 

Our results suggest that there is a weak but significant negative correlation between $v_{\mathrm{nt}}$ and $\theta$. One interpretation is that the $v_{\mathrm{nt}}$ is anisotropic with $v_{\mathrm{nt, \parallel}} > v_{\mathrm{nt, \perp}}$. However, this interpretation relies on the assumption that $v_{\mathrm{nt}}$ and its components are constant along the loop, so that the variations can be ascribed to the viewing angle and anisotropy. This is a common assumption made by other groups in previous studies of $v_{\mathrm{nt}}$ in active regions. 

In future measurements, we would like to better test this assumption by studying active regions with loops that have a nearly constant viewing angle and thereby  resolve whether the changes in $v_{\mathrm{nt}}$ are due to anisotropy or distance along the loop. There are a number of reasons that $v_{\mathrm{nt}}$ might vary along the loop. Wave amplitudes might be smaller near the footpoints due to the larger density there and grow in amplitude as the density decreases, or they might be larger near the footpoints close to the excitation source and decrease in amplitude due to dissipation as they travel up the loop, or the fraction of compressive waves that contribute to $v_{\mathrm{nt},\parallel}$ or Alfv\'en waves that contribute to $v_{\mathrm{nt},\perp}$ might also vary with $s$. 

Another limitation of these data is that the range of $\theta$ along the loop was limited to rather steep angles. For an active region near disk center, the line of sight is looking down onto the loop. As such, the vertical legs of the loop where the line of sight is nearly parallel to the line of sight are foreshortened and the legs are covered by only a few spatial pixels. Most of the field of view observes the region near the top of the loop where the line of sight is close to perpendicular. The anisotropy of $v_{\mathrm{nt}}$ would be better constrained by using observations that span a greater range of $\theta$. 

In order to help resolve these issues of variation along the loop and the limited dynamic range of angles, we plan to analyze an observation at the solar limb. At the limb, we expect to observe face-on and edge-on loops, much as was seen by \citet{Hara:ApJ:1999}, but with the advantage of using our improved magnetic field diagnostics to remove the ambiguity as to the actual magnetic field direction. 

\begin{acknowledgments}

This work was supported by the NASA Heliophysics Guest Investigator program grant NNM07AB07C. 

\end{acknowledgments}

\clearpage

\begin{thebibliography}{}
	\expandafter\ifx\csname natexlab\endcsname\relax\def\natexlab#1{#1}\fi
	\providecommand{\url}[1]{\href{#1}{#1}}
	\providecommand{\dodoi}[1]{doi:~\href{http://doi.org/#1}{\nolinkurl{#1}}}
	\providecommand{\doeprint}[1]{\href{http://ascl.net/#1}{\nolinkurl{http://ascl.net/#1}}}
	\providecommand{\doarXiv}[1]{\href{https://arxiv.org/abs/#1}{\nolinkurl{https://arxiv.org/abs/#1}}}
	
	\bibitem[{Asgari-Targhi {et~al.}(2021)Asgari-Targhi, Golub, Hahn, Karna, \&
		Savin}]{Asgari:ApJ:2021}
	Asgari-Targhi, M., Golub, L., Hahn, M., Karna, N., \& Savin, D.~W. 2021, ApJ,
	910, 113
	
	\bibitem[{Asgari-Targhi \& {van~Ballegooijen}(2012)}]{Asgari:ApJ:2012}
	Asgari-Targhi, M., \& {van~Ballegooijen}, A.~A. 2012, ApJ, 746, 81
	
	\bibitem[{Asgari-Targhi {et~al.}(2013)Asgari-Targhi, {van~Ballegooijen},
		Cranmer, \& DeLuca}]{Asgari:ApJ:2013}
	Asgari-Targhi, M., {van~Ballegooijen}, A.~A., Cranmer, S.~R., \& DeLuca, E.~E.
	2013, ApJ, 773, 111
	
	\bibitem[{Asgari-Targhi {et~al.}(2014)Asgari-Targhi, {van~Ballegooijen}, \&
		Imada}]{Asgari:ApJ:2014}
	Asgari-Targhi, M., {van~Ballegooijen}, A.~A., \& Imada, S. 2014, ApJ, 786, 28
	
	\bibitem[{Bobra {et~al.}(2008)Bobra, {van~Ballegooijen}, \& DeLuca}]{Bobra2008}
	Bobra, M.~G., {van~Ballegooijen}, A.~A., \& DeLuca, E.~E. 2008, ApJ, 672, 1209
	
	\bibitem[{Brooks \& Warren(2016)}]{Brooks:ApJ:2016}
	Brooks, D.~H., \& Warren, H.~P. 2016, ApJ, 820, 63
	
	\bibitem[{{Culhane} {et~al.}(2007){Culhane}, {Harra}, {James}, {Al-Janabi},
		{Bradley}, {Chaudry}, {Rees}, {Tandy}, {Thomas}, {Whillock}, {Winter},
		{Doschek}, {Korendyke}, {Brown}, {Myers}, {Mariska}, {Seely}, {Lang}, {Kent},
		{Shaughnessy}, {Young}, {Simnett}, {Castelli}, {Mahmoud}, {Mapson-Menard},
		{Probyn}, {Thomas}, {Davila}, {Dere}, {Windt}, {Shea}, {Hagood}, {Moye},
		{Hara}, {Watanabe}, {Matsuzaki}, {Kosugi}, {Hansteen}, \&
		{Wikstol}}]{Culhane:SolPhys:2007}
	{Culhane}, J.~L., {Harra}, L.~K., {James}, A.~M., {et~al.} 2007, Sol.\ Phys.,
	243, 19
	
	\bibitem[{{Del~Zanna} {et~al.}(2021){Del~Zanna}, Dere, Young, \&
		Landi}]{DelZanna:ApJ:2021}
	{Del~Zanna}, G., Dere, K.~P., Young, P.~R., \& Landi, E. 2021, ApJ, 909, 38
	
	\bibitem[{{Del~Zanna} \& Mason(2018)}]{DelZanna:LRSP:2018}
	{Del~Zanna}, G., \& Mason, H.~E. 2018, Living Reviews in Solar Physics, 15, 5
	
	\bibitem[{Dere(2022)}]{Dere:ApJ:2022}
	Dere, K. 2022, ApJ, 930, 86
	
	\bibitem[{Dere {et~al.}(1997)Dere, Landi, Mason, Fossi, \&
		Young}]{Dere:AA:1997}
	Dere, K.~P., Landi, E., Mason, H.~E., Fossi, B. C.~M., \& Young, P.~R. 1997,
	A\&AS, 125, 149
	
	\bibitem[{Feldman \& Landi(2008)}]{Feldman:PoP:2008}
	Feldman, U., \& Landi, E. 2008, Phys.\ Plasmas, 15, 056501
	
	\bibitem[{Hahn \& Savin(2013)}]{Hahn:ApJ:2013}
	Hahn, M., \& Savin, D.~W. 2013, ApJ, 763, 106
	
	\bibitem[{Hara \& Ichimoto(1999)}]{Hara:ApJ:1999}
	Hara, H., \& Ichimoto, K. 1999, ApJ, 513, 969
	
	\bibitem[{Hara {et~al.}(2011)Hara, Watanabe, Harra, Culhane, \&
		Young}]{Hara:ApJ:2011}
	Hara, H., Watanabe, T., Harra, L.~K., Culhane, J.~L., \& Young, P.~R. 2011,
	ApJ, 741, 107
	
	\bibitem[{Hara {et~al.}(2008)Hara, Watanabe, Harra, Culhane, Young, Mariska, \&
		Doschek}]{Hara:ApJ:2008}
	Hara, H., Watanabe, T., Harra, L.~K., {et~al.} 2008, ApJ, 678, 67
	
	\bibitem[{Klimchuk(2015)}]{Klimchuk:PTA:2015}
	Klimchuk, J.~A. 2015, Phil.\ Trans.\ A, 373, 20140256
	
	\bibitem[{{Lemen} {et~al.}(2012){Lemen}, {Title}, {Akin}, {Boerner}, {Chou},
		{Drake}, {Duncan}, {Edwards}, {Friedlaender}, {Heyman}, {Hurlburt}, {Katz},
		{Kushner}, {Levay}, {Lindgren}, {Mathur}, {McFeaters}, {Mitchell}, {Rehse},
		{Schrijver}, {Springer}, {Stern}, {Tarbell}, {Wuelser}, {Wolfson}, {Yanari},
		{Bookbinder}, {Cheimets}, {Caldwell}, {Deluca}, {Gates}, {Golub}, {Park},
		{Podgorski}, {Bush}, {Scherrer}, {Gummin}, {Smith}, {Auker}, {Jerram},
		{Pool}, {Soufli}, {Windt}, {Beardsley}, {Clapp}, {Lang}, \&
		{Waltham}}]{Lemen:SolPhys:2012}
	{Lemen}, J.~R., {Title}, A.~M., {Akin}, D.~J., {et~al.} 2012, Sol.\ Phys., 275,
	17
	
	\bibitem[{Rosner {et~al.}(1978)Rosner, Tucker, \& Vaiana}]{Rosner:ApJ:1978}
	Rosner, R., Tucker, W.~H., \& Vaiana, G.~S. 1978, ApJ, 220, 643
	
	\bibitem[{Savcheva \& {van~Ballegooijen}(2009)}]{Sav2009}
	Savcheva, A., \& {van~Ballegooijen}, A.~A. 2009, ApJ, 703, 1766
	
	\bibitem[{Scherrer {et~al.}(2012)Scherrer, Schou, Bush, Kosovichev, Bogart,
		Hoeksema, Liu, {Duvall~Jr.}, Zhao, Title, Schrijver, Tarbell, \&
		Tomczyk}]{Scherrer:SolPhys:2012}
	Scherrer, P.~H., Schou, J., Bush, R.~I., {et~al.} 2012, Sol.\ Phys., 275, 207
	
	\bibitem[{Su {et~al.}(2011)Su, Surges, {van~Ballegooijen}, DeLuca, \&
		Golub}]{Su2011}
	Su, Y., Surges, V., {van~Ballegooijen}, A., DeLuca, E., \& Golub, L. 2011, ApJ,
	734, 53
	
	\bibitem[{Su {et~al.}(2009{\natexlab{a}})Su, {van~Ballegooijen}, Lites, DeLuca,
		Golub, Grigis, Huang, \& Ji}]{Su2009a}
	Su, Y., {van~Ballegooijen}, A., Lites, B.~W., {et~al.} 2009{\natexlab{a}}, ApJ,
	691, 105
	
	\bibitem[{Su {et~al.}(2009{\natexlab{b}})Su, {van~Ballegooijen}, Schmieder,
		Berlicki, Guo, Golub, \& Huang}]{Su2009b}
	Su, Y.~N., {van~Ballegooijen}, A.~A., Schmieder, B., {et~al.}
	2009{\natexlab{b}}, ApJ, 704, 341
	
	\bibitem[{Testa {et~al.}(2016)Testa, {De~Pontieu}, \&
		Hansteen}]{Testa:ApJ:2016}
	Testa, P., {De~Pontieu}, B., \& Hansteen, V. 2016, ApJ, 827, 99
	
	\bibitem[{{van~Ballegooijen}(2004)}]{vanB2004}
	{van~Ballegooijen}, A.~A. 2004, ApJ, 612, 519
	
	\bibitem[{{van~Ballegooijen} {et~al.}(2000){van~Ballegooijen}, Priest, \&
		Mackay}]{vanB2000}
	{van~Ballegooijen}, A.~A., Priest, E.~R., \& Mackay, D.~H. 2000, ApJ, 539
	
	\bibitem[{Warren {et~al.}(2014)Warren, Ugarte-Urra, \&
		Landi}]{Warren:ApJS:2014}
	Warren, H.~P., Ugarte-Urra, I., \& Landi, E. 2014, ApJS, 213, 11
	
	\bibitem[{Yang {et~al.}(1986)Yang, Sturrock, \& Antiochos}]{Yang1986}
	Yang, W.~H., Sturrock, P.~A., \& Antiochos, S.~K. 1986, ApJ, 309, 383
	
	\bibitem[{Young(2011)}]{Young:EIS:2011}
	Young, P. 2011
	
\end{thebibliography}
\end{document}